\documentclass[twocolumn,pra,showpacs,superscriptaddress]{revtex4-1}

\usepackage{amsmath}
\usepackage{amsthm}
\usepackage{xcolor}
\usepackage{pgf,tikz}

\DeclareMathOperator {\Tr}{Tr}

\newtheorem{lemma}{Lemma}
\newtheorem{proposition}{Proposition}

\begin{document}

\title{Optimal transfer of an unknown state via a bipartite operation}

\author{Yang Liu} 

\affiliation{Beijing National Laboratory for
  Condensed Matter Physics, and Institute of Physics, Chinese Academy
  of Sciences, Beijing 100190, China}

\author{Yu Guo}

\affiliation{Beijing National Laboratory for Condensed Matter Physics,
  and Institute of Physics, Chinese Academy of Sciences, Beijing
  100190, China}

\affiliation{School of Physics and Electronic Science, Changsha
  University of Science and Technology, Changsha 410114, China}

\author{D.~L. Zhou}

\affiliation{Beijing National Laboratory for Condensed Matter Physics,
  and Institute of Physics, Chinese Academy of Sciences, Beijing
  100190, China}

\begin{abstract}
  A fundamental task in quantum information science is to transfer an
  unknown state from particle $A$ to particle $B$ (often in remote
  space locations) by using a bipartite quantum operation
  $\mathcal{E}^{AB}$. We suggest the power of $\mathcal{E}^{AB}$ for
  quantum state transfer (QST) to be the maximal average probability
  of QST over the initial states of particle $B$ and the
  identifications of the state vectors between $A$ and $B$. We find
  the QST power of a bipartite quantum operations satisfies four
  desired properties between two $d$-dimensional Hilbert spaces. When
  $A$ and $B$ are qubits, the analytical expressions of the QST power
  is given. In particular, we obtain the exact results of the QST
  power for a general two-qubit unitary transformation.
\end{abstract}


\maketitle

\section{Introduction}

A fundamental task in quantum information science is to transfer an
unknown internal quantum state of a particle from one location
$\mathcal{A}$ to another location $\mathcal{B}$. A direct method is to
mechanically move the particle from $\mathcal{A}$ to $\mathcal{B}$
while keeping the internal state invariant. A more sophisticated way
is quantum state teleportation ~\cite{PhysRevLett.70.1895}, where the
unknown state is teleported with the aid of a pair of particles in a
Bell state and $2$ bits of classical communications. The third way is
to transfer the state via a two-particle quantum operation
$\mathcal{E}^{AB}$, which can be realized by linking two nodes A and B
to a quantum network, e.g., a quantum wire (a one-dimensional chain of
particles with interactions) ~\cite{PhysRevLett.78.3221,
  PhysRevLett.91.207901, PhysRevLett.92.187902}. Here the node A,
located in $\mathcal{A}$, is the particle with the unknown state to be
transferred, and the node B, located in $\mathcal{B}$, is the particle
as the state receiver.

The aim of Refs.~\cite{PhysRevLett.91.207901, PhysRevLett.92.187902,
  PhysRevA.71.032309, PhysRevLett.98.010501, PhysRevLett.101.230502,
  PhysRevLett.106.040505, PhysRevLett.109.050502} is to achieve
perfect quantum state transfer by optimizing the quantum network. In
addition, the capacity of quantum state transfer to characterize the
non-locality of a bipartite unitary transformation is studied in
Refs. ~\cite{PhysRevA.64.032302, Hammerer2002}.  Here we will solve
another related question: For a given two-particle quantum operation
$\mathcal{E}^{AB}$, what is the maximal average probability for
quantum state transfer? This maximal probability reflects the power of
quantum state transfer of the bipartite operation $\mathcal{E}^{AB}$.

The approach we will adopt is similar to that in the power of
entanglement generation for a local unitary gate
~\cite{PhysRevA.62.030301, PhysRevA.63.062309,
  PhysRevA.67.052301}. Here we want to emphasize that the bipartite
quantum operations, including non-unitary gates, are necessary
to be considered for quantum state transfer.

The article is organized as follows. In Sec. II, we introduce the
basic formula of the power of quantum state transfer, and four basic
properties for the QST power are proved. In Sec. III, we give the
analytical results of two-qubit operations. In particular, an
exact result of the QST power for any two-qubit unitary transformation
is given. Finally we present some discussions and a brief summary.   

\section{General results}

\subsection{The power of QST}

In this section, we will give a proper quantity to measure the power
of QST for a bipartite quantum operation.

\begin{figure}[htp]
\begin{tikzpicture}
\draw (1,1) rectangle (3,2);
\path (2,1.5) node (na) {$\mathcal{E}^{AB}$};
\draw[->] (1.4,0.5)--(1.4,1.0);
\draw[->] (1.4,2.0)--(1.4,2.5);
\draw[->] (2.6,0.5)--(2.6,1.0);
\draw[->] (2.6,2.0)--(2.6,2.5);
\node at (1.4,0.2) {$\vert\psi^A\rangle$};
\node at (2.6,0.2) {$\vert\xi^B\rangle$};
\node at (2.6,2.8) {$S^B\vert\psi^B\rangle$};
\end{tikzpicture}
\caption{The process of quantum state transfer using a bipartite
  quantum operation $\mathcal{E}^{AB}$. We maximize the probability of
  QST over particle B's initial states $\vert\xi^B\rangle$ and the
  unitary transformations $S^B$.}
\label{fig:1}
\end{figure}
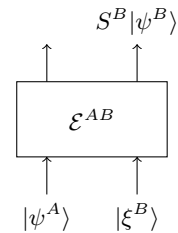

We consider two particles $A$ and $B$, whose Hilbert space is
$\mathcal{H}^{AB}=\mathcal{H}^{A}\otimes\mathcal{H}^{B}$ with
$\dim\mathcal{H}^{A}=\dim\mathcal{H}^{B}=d$. In other words, particles
A and B are two qudits. Initially particle $A$ is prepared in an
unknown state $\vert\psi^{A}\rangle$, and particle $B$ is in some
given state $\vert\xi^{B}\rangle$. For convenience, we take
$\vert\psi^{A}\rangle=R\vert{0^{A}}\rangle$ with $\vert{0^{A}}\rangle$
being any given state and $R\in SU(d)$. In general, there are many
choices of $R$ for given $\vert\psi^{A}\rangle$ and
$\vert{0^{A}}\rangle$, which does not affect the following
formulations. After performing a bipartite quantum operation
$\mathcal{E}^{AB}$, we need to estimate to which degree the unknown
state $\vert\psi^{A}\rangle$ being transferred to partilce $B$. This
process of QST is depicted in Fig. \ref{fig:1}.

For a given bipartite quantum operations, we may improve the QST by
controlling two elements. On one hand, the QST power of
$\mathcal{E}^{AB}$ depends on the input state of particle $B$
$\vert\xi^{B}\rangle$.  We can improve the QST by preparing particle
$B$ in a suitable initial state. On the other hand, notice that if the
final state is $S^{B}\left|\psi^{B}\right\rangle$, where $S^{B}$ is a
unitary transformation on particle $B$ independent of the unknown state
$\vert\psi^{A}\rangle$, the unkown state will be regarded as being
perfectly transferred. Different choices of $S^B$ implies different
identifications of the bases between $\mathcal{H}^A$ and
$\mathcal{H}^B$. Hence we can improve the QST of $\mathcal{E}^{AB}$ by
adjusting $S^B$.

To give a measure to characterize
the power of QST for $\mathcal{E}^{AB}$, we need to optimize over
$\vert\xi^{B}\rangle$ and $S^{B}$. Therefore the QST power for a
quantum operation $\mathcal{E}^{AB}$ is defined as
\begin{equation}
  \mathcal{P}_{QST}\left(\mathcal{E}^{AB}\right)  = 
  \max_{S^{B}, \left|\xi^{B}\right\rangle} 
  \int d\mu\left(R\right) P(R;\mathcal{E}^{AB};S^{B},\vert\xi^B\rangle)
\label{def:qst}
\end{equation}
where
\begin{equation}
  P(R;\mathcal{E}^{AB};S^B,\vert\xi^B\rangle) = \Tr\nolimits
  \left(\mathcal{E}^{AB}
    \mathcal{R}^{A}\left(P_{0}^{A}P_{\xi}^{B}\right)
    \mathcal{S}^{B}\mathcal{R}^{B} (P_{0}^{B})\right)
\label{eq:prob}
\end{equation}
with $P_{0}^{A}=\vert0{}^{A}\rangle\langle0{}^{A}\vert$,
$P_{\xi}^{B}=\vert\xi{}^{B}\rangle\langle\xi{}^{B}\vert$,
$P_{0}^{B}=\vert0{}^{B}\rangle\langle0{}^{B}\vert$, $\mathcal{R}$ and
$\mathcal{S}$ are the local quantum operations corresponding to the
unitary transformations $R$ and $S$, and $d\mu\left(R\right)$ being
the Haar measure on $SU\left(d\right)$.  Because the Harr measure is
right-invariant, the power of QST is independent of the choice of
$\left|0\right\rangle$. Since we have no reasons to give different
probabilities to two sets of states connecting by a unitary
transformation in the average over the unknown states, the Harr
measure is a natural choice for the average.

For a given initial state $R^A\vert{0}\rangle\otimes\vert\xi^B\rangle$
and a given identification between $\mathcal{H}^A$ and $\mathcal{H}^B$
specified by $S^B$, $P(R;\mathcal{E}^{AB};S^B,\vert\xi^B\rangle$ is
the probability of QST for the state $R\vert{0}\rangle$ by
$\mathcal{E}^{AB}$. Further more, the power of QST for
$\mathcal{E}^{AB}$ is the maximal average probability for particle $B$
in the unkown state after the action of $\mathcal{E}^{AB}$.

\subsection{Properties of the QST power}

In the above subsection, we give a definition of the QST power for a
bipartite quantum operation, and give its physical
interpretation. Here we will prove that the QST power
$\mathcal{P}_{QST}\left(\mathcal{E}^{AB}\right)$ has the following
desired properties.

\textbf{Property (i).} The QST power is invariant under local unitary
transformations.

Let $X^{A}$, $Y^{B}$, $U^{A}$, $V^{B}$ be local unitary
transformations, and let $\mathcal{X}^{A}$, $\mathcal{Y}^{B}$,
$\mathcal{U}^{A}$, $\mathcal{V}^{B}$ be the corresponding local
quantum operations.  Then
\begin{eqnarray}
  &  & \mathcal{P}_{QST}\left( \mathcal{X}^{A} \mathcal{Y}^{B}
    \mathcal{E}^{AB} \mathcal{U}^{A} \mathcal{V}^{B} \right) \nonumber\\
  & = & \max_{S^{B},\left|\xi^{B}\right\rangle } \int d\mu(R)
  P(R;\mathcal{X}^{A} \mathcal{Y}^{B} \mathcal{E}^{AB} \mathcal{U}^{A}
  \mathcal{V}^{B};S^B,\vert\xi^B \rangle) \nonumber\\
  & = & \max_{S^{B}, \left|\xi^{B}\right\rangle} \int d\mu(R)
  P(UR;\mathcal{E}^{AB};Y^{B\dagger}S^BV^{B\dagger},V^B\vert\xi^B\rangle)
  \nonumber\\
  & = & \max_{S^{B},\left|\xi^{B}\right\rangle }\int d\mu(R)
  P(R;\mathcal{E}^{AB};S^B,\vert\xi^B\rangle) \nonumber\\
  & = & \mathcal{P}_{QST}\left(\mathcal{E}^{AB}\right),
\label{eq:pro1}
\end{eqnarray}
where we have used Eq. (\ref{eq:inva}) in the third line of the above
equation.

Property (i) shows the power of a bipartite quantum operation
characterizes its nonlocal property ~\cite{PhysRevA.67.052301}.

\textbf{Property (ii).} The range of the QST power is in the period
$\left[\frac{1}{d},1\right]$, i.e.,
\begin{equation}
  \frac {1} {d} \le \mathcal{P}_{QST} \left(\mathcal{E}^{AB}\right)
  \le 1.
\label{eq:pro2}
\end{equation}

This can be proved as follows. Because
\begin{eqnarray*}
  &  & \mathcal{P}_{QST} \left(\mathcal{E}^{AB}\right)\\
  & = & \max_{S^{B},\left|\xi^{B}\right\rangle } \int d\mu
  \left(R\right)  P(R;\mathcal{E}^{AB};S^B,\vert\xi^B\rangle)\\
  & \le & \max_{S^{B},\left|\xi^{B}\right\rangle } \int
  d\mu\left(R\right) 1 = 1.
\end{eqnarray*}
In addition, if
$\mathcal{P}_{QST}\left(\mathcal{E}^{AB}\right)<\frac{1}{d}$, then
\begin{eqnarray*}
  &  & \int d\mu\left(S\right) \int d\mu\left(R\right) 
  P(R;\mathcal{E}^{AB};S^B,\vert\xi^B\rangle) \\
  & \le & \int d\mu \left(S\right) \mathcal{P}_{QST}
  \left(\mathcal{E}^{AB}\right) < \frac {1} {d}.
\end{eqnarray*}
However, 
\begin{eqnarray*}
  &  & \int d\mu\left(S\right) \int d\mu\left(R\right)
  P(R;\mathcal{E}^{AB};S^B,\vert\xi^B\rangle)\\
  & = & \int d\mu\left(R\right) \Tr\nolimits \left( \mathcal{E}^{AB}
    \left(R^{A}P_{0}^{A}R^{A\dagger}P_{\xi}^{B}\right) \frac {I^{B}}
    {d}  \right)\\
  & = & \int d\mu\left(R\right) \frac {1} {d} = \frac {1} {d}.
\end{eqnarray*}
In the second line, we used the lemma (\ref{lemma}). This result
contradicts with the above inequality. Therefore
$\frac{1}{d}\le\mathcal{P}_{QST}\left(\mathcal{E}^{AB}\right)$.

Notice that the lower bound $1/d$ is the same as in the case of
transferring a classical discrete
variable with $d$ different states. 

\textbf{Property (iii).} The QST power of a local operation
$\mathcal{E}^{A}\otimes\mathcal{E}^{B}$ is $\frac{1}{d}$.
\begin{eqnarray*}
  &  & \mathcal{P}_{QST} \left( \mathcal{E}^{A} \otimes
    \mathcal{E}^{B}  \right)\\
  & = & \max_{S^{B},\left|\xi^{B}\right\rangle } \int d\mu\left(R\right)
  P(R;\mathcal{E}^{A}\mathcal{E}^{B};S^B,\vert\xi^B\rangle) \\
  & = & \max_{S^{B},\left|\xi^{B}\right\rangle } \int
  d\mu\left(R\right) \Tr\nolimits \left(\mathcal{E}^{B}
    \left(P_{\xi}^{B}\right) S^{B} R^{B} P_{0}^{B} R^{B\dagger}
    S^{B\dagger} \right)\\
  & = & \max_{\left|\xi^{B}\right\rangle } \int d\mu\left(R\right)
  \Tr\nolimits \left(\mathcal{E}^{B} \left(P_{\xi}^{B}\right) R^{B}
    P_{0}^{B} R^{B\dagger} \right)\\
  & = & \max_{\left|\xi^{B}\right\rangle } \Tr\nolimits
  \left( \mathcal{E}^{B} \left(P_{\xi}^{B}\right)
    \frac {I^{B}} {d} \right)\\
  & = & \frac{1}{d}.
\end{eqnarray*}
In the fifth line of the above equation, we use the lemma
(\ref{lemma}).

As expected, a local operation has the lowest power in transferring an
unknown state.

\textbf{Property (iv).} The QST power of the swapping gate $SWAP^{AB}$
is $1$.
\begin{eqnarray*}
  &  & \mathcal{P}_{QST} \left(SWAP^{AB}\right)\\
  & = & \max_{S^{B},\left|\xi^{B}\right\rangle } \int d \mu\left(R\right)
  P(R;{SWAP}^{AB};S^B,\vert\xi^B\rangle) \\
  & = & \max_{S^{B},\left|\xi^{B}\right\rangle } \int
  d\mu\left(R\right) \Tr\nolimits \left(R^{B}P_{0}^{B}R^{B\dagger}
    P_{\xi}^{A} S^{B} R^{B} P_{0}^{B} R^{B\dagger} S^{B\dagger} \right)\\
  & = & \max_{S^{B}} \int
  d\mu\left(R\right) \Tr\nolimits \left( R^{B} P_{0}^{B} R^{B\dagger}
    S^{B} R^{B} P_{0}^{B} R^{B\dagger} S^{B\dagger} \right)\\
  & = & \int d\mu\left(R\right) \Tr\nolimits \left(
    R^{B} P_{0}^{B} R^{B\dagger} R^{B} P_{0}^{B} R^{B\dagger} \right)\\
  & = & 1.
\end{eqnarray*}
This result is reasonable because the unknown state is swapped, i.e.,
perfectly transferred.

\section{Analytical results of the qubit case}

Since the QST power is defined as an optimization problem over a state
and a unitary transformation in a $d$-dimensional Hilbert space, the
explicit calculations of the QST power of $\mathcal{E}^{AB}$ when A
and B are two qudits, in general, are complex. In this section, we
will give an analytical result on the QST power for an arbitrary
two-qubit quantum operation $\mathcal{E}^{AB}$, which makes the
numerical calculation of the QST power becomes accessible. In
particular, we further obtain the exact result of the QST power for
any two-qubit unitary transformation.

For the qubit case, let
$P_0=\frac{I+\sigma_z}{2}$. $\forall{R}\in{SU(2)}$, we can find a
coordinate frame $\{\vec{R}_x,\vec{R}_y,\vec{R}_z\}$ to characterize
it. The base vector of the coordinate frame is defined by
$R\sigma_{n}R^{\dagger}=\vec{R}_{n}\cdot\vec{\sigma}$. Then the
component of the bases vector
$R_{n}^{m}=\frac{\Tr\left(R\sigma_{n}R^{\dagger}\sigma_{m}\right)}{2}$. The
initial state of particle B
$P_{\xi}^{B}=\frac{I^{B}+\vec{T}\cdot\vec{\sigma}^{B}}{2}$, where
$\vec{T}$ is the Bloch vector for the state $\vert\xi^B\rangle$.

The QST power of $\mathcal{E}^{AB}$ is
\begin{equation}
\mathcal{P}_{QST} \left(\mathcal{E}^{AB}\right) = 
 \frac {1} {2} + \frac {\max\limits_{S,T} \left( \sum \limits_{l,n}
     \mathcal{E}_{0n}^{l} S_{l}^{n} + \sum \limits_{l,m,n}
     \mathcal{E}_{mn}^{l} T^{m} S_{l}^{n} \right)} {24},
\label{eq:qstp2}
\end{equation}
where
\begin{eqnarray*}
  \mathcal{E}_{0n}^{l} & = & \Tr \left( \mathcal{E}^{AB} \left(
      \sigma_{l}^{A}I^{B} \right) \sigma_{n}^{B} \right),\\
  \mathcal{E}_{mn}^{l} & = & \Tr\left( \mathcal{E}^{AB}
    \left(\sigma_{l}^{A} \sigma_{m}^{B} \right) \sigma_{n}^{B} \right),
\end{eqnarray*}
and $\{S_n\}$ is the basis vectors of the coordinate frame defined by
the unitary transformation $S$.

To derive Eq. (\ref{eq:qstp2}), we used the following Harr average
values on $SU(2)$:
\begin{eqnarray*}
  \left\langle R_{z}^{m} \right\rangle  & = & 0,\\
  \left\langle R_{z}^{m} R_{z}^{n} \right\rangle  & = & \frac {1} {3}
  \delta_{mn},
\end{eqnarray*}
 whose detailed proofs are given in the appendix B. 

 Let us demonstrate the power of Eq. (\ref{eq:qstp2}) with calculating
 the QST power of the CNOT gate. A direct calculation gives
\begin{eqnarray*}
\mathcal{P}_{QST} \left(CNOT^{AB}\right)
 = \frac {1} {2} + \frac {\max_{S,T} \left( T^{x} S_{z}^{x} +
     T^{y} S_{z}^{y} \right)} {6}
 =  \frac {2} {3}.
\end{eqnarray*}

It is worthy to point out that Eq. (\ref{eq:qstp2}) can be used as the
foundation for numerical calculations of the QST power for arbitrary
two-qubit quantum operation. For example, it may find applications in
the process of QST along a quantum wire ~\cite{PhysRevLett.91.207901,
  PhysRevLett.92.187902, PhysRevA.71.032309}.

\subsection{Exact result on QST power for arbitrary two-qubit unitary
  transformations}

In this subsection, we will apply Eq. (\ref{eq:qstp2}) to the case
when $\mathcal{E}^{AB}$ is a two-qubit unitary transformation. In this
case, the exact result of the QST power will be obtained.

First notice that a general unitary transformation for two qubits can
be written as
\[
U^{AB} = U^{A} U^{B} U_{d}^{AB} V^{B} V^{A},
\]
where
\[
U_{d}^{AB} = e^{-\frac {i} {2} \left(\sum_{m} d_{m} \sigma_{m}^{A}
    \sigma_{m}^{B} \right)}
\]
with $\left|d_{z}\right|\le d_{y}\le d_{x}\le\frac{\pi}{2}$
~\cite{PhysRevA.63.032308, PhysRevA.63.062309}. Because
the QST power is invariant under local unitary transformations, it is
sufficient to study the unitary transformation $U_{d}^{AB}$. 

Through a complex but direct calculation, we arrives at
\begin{eqnarray*}
  \mathcal{P}_{QST} \left(U^{AB}\right) 
  =  \frac {1} {2} + \frac {\max_{S,T}f} {6},
\end{eqnarray*}
where 
\begin{eqnarray*}
f && =  \sin d_{y} \sin d_{z} S_{x}^{x} + \sin d_{z} \sin
d_{x} S_{y}^{y} + \sin d_{x} \sin d_{y} S_{z}^{z}\\
 && +  \cos d_{y} T^{y} \sin d_{x} S_{x}^{z} + \cos d_{z} T^{z} \sin
 d_{y} S_{y}^{x} + \cos d_{x} T^{x} \sin d_{z} S_{z}^{y}\\
 && - \cos d_{z} T^{z} \sin d_{x} S_{x}^{y} - \cos d_{x} T^{x} \sin
 d_{y} S_{y}^{z} - \cos d_{y} T^{y} \sin d_{z} S_{z}^{x}.
\end{eqnarray*}

Notice that 
\[
f \left( -\vec{S}_{x},-\vec{S}_{y},\vec{S}_{z},-\vec{T} \vert d_{z} \le
  0 \right) = f \left( \vec{S}_{x},\vec{S}_{y},\vec{S}_{z},\vec{T}
  \vert d_{z} \ge 0 \right).
\]
Hence 
$\max{f}\left(d_{z}\right)=\max{f}\left(-d_{z}\right)$.
Therefore we only need to study the case when $d_{z}\ge{0}$.

We obtain the exact result on the maximization of $f$:
\begin{equation}
  \max_{S,T} f = \sin d_x +\sin d_y +\sin d_x \sin d_y.
\label{eq:maxf}
\end{equation}
The proof of Eq. (\ref{eq:maxf}) can be found in the appendix
$C$. Therefore the power of a two-qubit unitary transformation is
\begin{equation}
  \mathcal{P}_{QST} \left(U^{AB}\right) 
  =  \frac {1} {2} + \frac {\sin d_x + \sin d_y + \sin d_x \sin d_y} {6}.
\label{eq:qstp2u}
\end{equation}

A remarkable feature in the QST power of $U^{AB}$ is that it is
independent of the parameter $d_z$. For the CNOT gate, $d_x=\pi/2$ and
$d_y=d_z=0$, so its QST power is $2/3$, which is the same as
calculated above. To make a perfect QST, we require that
$d_x=d_y=\pi/2$. To make the QST power lowest, we get $d_x=d_y=0$,
where $U^{AB}$ becomes a local unitary transformation. 

\section{Discussions and summary}

In Refs. ~\cite{PhysRevA.64.032302, Hammerer2002}, the nonlocal
properties of a bipartite gate are classified according their
capacities in transmitting classical or quantum bits of
information. Here we suggest the QST power to characterize the
capacity to transmit quantum state.  In parallel, our method can be
generalized to classical state transfer or quantum state swapping.

Technically, we prove that the QST power of a local bipartite quantum
operation is $1/d$. However we don't know whether the QST power of a
bipartite quantum operation is $1/d$ implies that the bipartite
quantum operation is local. In addition, we give a lengthy proof of
the maximization, Eq. (\ref{eq:maxf}), in Appendix C. Does there
exist some simpler proof of Eq. (\ref{eq:maxf})?

In summary, we suggest the QST power of a bipartite quantum operation
as the maximal average probability of QST using the bipartite quantum
operation. Four basic properties of the QST power of a bipartite
quantum operation are proved. Then we obtain the analytical result of
the QST power for any two-qubit quantum operation, which may be used
as the foundation for numerical calculations of the QST power. The
exact result of the QST power for arbitrary two-qubit unitary
transformation is obtained. We hope that our work present an
alternative method to characterize the non-locality of a bipartite
quantum operation.

\begin{acknowledgments}
  We thank S.L. Luo, Z.W. Zhou, and C.P. Sun for helpful
  discussions.  This work is supported by NSF of China (Grant
  Nos. 10975181 and 11175247) and NKBRSF of China (Grant
  No. 2012CB922104).
\end{acknowledgments}

\begin{appendix}
  
\section{Harr measure}
Notice that the Harr measure satisfies two useful properties
~\cite{Rudi91}:

i) It is normalized.
\begin{equation} 
  \int d\mu(R) 1 = 1.
\label{eq:norm}
\end{equation}

ii) It is left-invariant and right-invariant. 
$\forall S\in SU(d)$,
\begin{equation}
  \int d\mu(R) f(R) = \int d\mu(R) f(S R) = \int d\mu(R) f(RS). 
\label{eq:inva}
\end{equation}

 \begin{lemma}
$\forall\vert0\rangle\in\mathcal{H}$ and $R\in{SU(d)}$, we have
\begin{equation}
  \int d\mu(R) R P_0 R^\dagger = \frac {I} {d}.
\label{lemma}
\end{equation}
\end{lemma}
This can be proved as follows. Firstly we take a complete normal
orthogonal bases of $\mathcal{H}$, denoted as
$\{\vert{n}\rangle,\;n\in\{0,1,\cdots,d-1\}\}$. Because $\forall{n}$
there exists a unitary transformation $S_n$ such that
$\vert{n}\rangle=S_n\vert{0}\rangle$, Eq. (\ref{eq:inva}) gives
$\int{d}\mu(R)R{P_n}R^\dagger=\int{d}\mu(R)R{P_0}R^\dagger$.
Therefore
$\int{d}\mu(R)R{P_0}R^\dagger=\int{d}\mu(R)R\frac{\sum_n{P_n}}{d}R^\dagger=\frac{I}{d}$.

\section{Harr average on $SU(2)$}

\begin{proposition}
$\forall m,n\in\{x,y,z\}$, we have
\begin{eqnarray}
  \langle R_z^m \rangle &=& 0, \nonumber\\
  \langle R_z^m R_z^n\rangle &=& \frac {1} {3} \delta_{mn}.\nonumber
\end{eqnarray}
\end{proposition}

We can prove the above result by a direct calculation. Here we present
an alternative approach as follows.

$\forall m\in\{x,y,z\}$, $\exists n\neq m$, 
\begin{eqnarray*}
  \left\langle R_{z}^{m}\right\rangle  & = & \int d\mu\left(R\right) 
  \frac {\Tr\left(R\sigma_{z}R^{\dagger} \sigma_{m} \right)} {2}\\
  & = & \int d\mu\left(R\right) \frac{\Tr \left( \sigma^{n} R
      \sigma_{z} R^{\dagger} \sigma_{n}^{\dagger} \sigma_{m} \right)} {2}\\
  & = & -\int d\mu\left(R\right) \frac {\Tr \left(R \sigma_{z}
      R^{\dagger} \sigma_{m} \right)} {2}\\
  & = & -\left\langle R_{z}^{m} \right\rangle .
\end{eqnarray*}
Therefore $\left\langle R_{z}^{m}\right\rangle=0$.

When $m\ne{n}$,
\begin{eqnarray*}
  &&\left\langle R_{z}^{m} R_{z}^{n} \right\rangle  =  \int d\mu
   \left(R\right) \frac {\Tr\left(R^{A} \sigma_{z}^{A} R^{A\dagger}
      \sigma_{m}^{A} R^{B} \sigma_{z}^{B} R^{B\dagger} \sigma_{n}^{B}
    \right)} {4}\\
  && =  \int d\mu\left(R\right) \frac {\Tr \left(\sigma_{m}^{A} R^{A}
      \sigma_{z}^{A} R^{A\dagger} \sigma_{m}^{A\dagger} \sigma_{m}^{A}
      \sigma_{m}^{B}R^{B} \sigma_{z}^{B} R^{B\dagger}
      \sigma_{m}^{B\dagger} \sigma_{n}^{B\dagger} \right)} {4}\\
  && =  -\int d\mu\left(R\right) \frac {\Tr \left(R^{A} \sigma_{z}^{A}
      R^{A\dagger} \sigma_{m}^{A} R^{B} \sigma_{z}^{B} R^{B\dagger}
      \sigma_{n}^{B} \right)} {4}\\
  && =  -\left\langle R_{z}^{m} R_{z}^{n} \right\rangle .
\end{eqnarray*}
Hence, $\left\langle R_{z}^{m}R_{z}^{n}\right\rangle =0$ if $m\neq n$.

When $m\neq n$, $\exists$ a unitary transformation $H$, such that
$H^{\dagger}\sigma_{m}H=\sigma_{n}$. Hence 
\begin{eqnarray*}
  &&\left\langle R_{z}^{m} R_{z}^{m} \right\rangle  =  \int d\mu
  \left(R\right) \frac {\Tr\left(R^{A} \sigma_{z}^{A} R^{A\dagger}
      \sigma_{m}^{A} R^{B} \sigma_{z}^{B} R^{B\dagger} \sigma_{m}^{B}
    \right)} {4}\\
  && =  \int d\mu \left(R\right) \frac {\Tr \left(H^{A} R^{A}
      \sigma_{z}^{A} R^{A\dagger} H^{A\dagger} \sigma_{m}^{A} H^{B}
      R^{B} \sigma_{z}^{B} R^{B\dagger} H^{B\dagger}
      \sigma_{m}^{B\dagger} \right)}  {4}\\
  && =  \int d\mu\left(R\right) \frac {\Tr \left(R^{A} \sigma_{z}^{A}
      R^{A\dagger} \sigma_{n}^{A} R^{B} \sigma_{z}^{B} R^{B\dagger}
      \sigma_{n}^{B} \right)} {4}\\
  && =  \left \langle R_{z}^{n} R_{z}^{n} \right\rangle .
\end{eqnarray*}
Because 
$\sum_{m}\left(R_{z}^{m}\right)^{2}=1$,
we have
$\left\langle{R_{z}^{m}}R_{z}^{m}\right\rangle=\frac{1}{3}$.

\section{Maximization of the function $f$}

In this appendix, we will prove Eq. (\ref{eq:maxf}). First, we divide
$f$ into two parts:
\[
f = f_1 + f_2,
\]
where
\begin{eqnarray*}
  f_{1} & = & \sin{d_{y}} \sin{d_{z}} S_{x}^{x} + \sin{d_{z}}
  \sin{d_{x}} S_{y}^{y} + \sin{d_{x}} \sin{d_{y}} S_{z}^{z},\\
  f_{2} & = & \lambda a +\mu b + \gamma c + \eta g + \nu p + \xi q
\end{eqnarray*}
with $\lambda=T^{z}S_{x}^{y}$, $\mu=-T^{z}S_{y}^{x}$,
$\gamma=-T^{y}S_{x}^{z}$, $\eta=T^{y}S_{z}^{x}$, $\nu=T^{x}S_{y}^{z}$,
$\xi=-T^{x}S_{z}^{y}$, and $a=\cos{d_{z}}\sin{d_{x}}$,
$b=\cos{d_{z}}\sin{d_{y}}$, $c=\cos{d_y}\sin{d_{x}}$,
$g=\cos{d_y}\sin{d_{z}}$, $p=\cos{d_x}\sin{d_y}$,
$q=\cos{d_x}\sin{d_{z}}$.

Because $\vec{T}$ is a unit vector, and
$\{\vec{S}_x,\vec{S}_y,\vec{S}_z\}$ is a right-handed coordinate
frame, they are parameterized as follows.
\begin{eqnarray}
  T^{x} & = & \sin\alpha \cos\beta,\label{eq:1}\\
  T^{y} & = & \sin\alpha \sin\beta,\label{eq:2}\\
  T^{z} & = & \cos\alpha,\label{eq:3}\\
  S_{x}^{x} & = & (\sin^2\phi + \cos^{2}\theta \cos^{2}\phi) \cos\omega
  + \cos^{2}\phi \sin^{2}\theta,\label{eq:4}\\
  S_{x}^{y} & = & \cos\phi \sin\phi+\cos\theta \sin\omega - \cos\phi
  \cos^{2}\theta \sin\phi \nonumber \\
  &  & - \cos\omega \cos\phi \sin\phi + \cos\omega \cos\phi
  \cos^{2}\theta \sin\phi, \label{eq:5}\\
  S_{x}^{z} & = & \sin\theta (\cos\phi \cos\theta (1 - \cos\omega) -
  \sin\omega \sin\phi), \label{eq:6}\\
  S_{y}^{x} & = & \cos\phi \sin\phi - \cos\theta \sin\omega - \cos\phi
  \cos^{2}\theta \sin\phi \nonumber \\
  &  & - \cos\omega \cos\phi \sin\phi + \cos\omega \cos\phi
  \cos^{2}\theta \sin\phi, \label{eq:7}\\
  S_{y}^{y} & = & (\cos^{2}\phi + \cos^2\theta \sin^2\phi) \cos\omega +
  \sin^{2}\theta \sin^{2}\phi, \label{eq:8}\\
  S_{y}^{z} & = & \sin\theta (\cos\phi \sin\omega + (1 - \cos\omega)
  \cos\theta \sin\phi), \label{eq:9}\\
  S_{z}^{x} & = & \sin\theta (\sin\omega \sin\phi + (1-\cos\omega)
  \cos\phi \cos\theta), \label{eq:10}\\
  S_{z}^{y} & = & \sin\theta ((1 - \cos\omega) \cos\theta \sin\phi -
  \cos\phi \sin\omega), \label{eq:11}\\
  S_{z}^{z} & = & \cos^{2}\theta + \cos\omega \sin^{2}\theta,\label{eq:12}
\end{eqnarray}
where $0\leq\alpha,\theta,\omega\leq\pi$, $0\leq\beta,\phi\leq2\pi$.

\subsection{Analysis of $f_{1}$}

Using Eq.(\ref{eq:4}), Eq.(\ref{eq:8}), Eq.(\ref{eq:12}), we can
rewrite $f_{1}$ as
\begin{eqnarray*}
f_{1} & = & \sin{d_y} \sin{d_{z}} S_{x}^{x} + \sin{d_{z}} \sin{d_x}
S_{y}^{y} + \sin{d_x} \sin{d_y} S_{z}^{z}\\
 & = & \left(1 - \cos\omega \right) M \cos^{2}\theta + \sin{d_x}
 \sin{d_y} \cos\omega\\
 &  & + \sin{d_y} \sin{d_{z}} \left(\cos^{2}\phi + \cos\omega
   \sin^{2}\phi \right)\\
 &  & + \sin{d_x} \sin{d_{z}} \left(\sin^{2}\phi + \cos\omega
   \cos^{2}\phi \right).
\end{eqnarray*}
Because 
\begin{eqnarray*}
  M & = & \sin{d_x} \sin{d_y} - \sin{d_y} \sin{d_{z}} \cos^{2}\phi -
  \sin{d_x} \sin{d_{z}} \sin^{2}\phi\\
  & \geq & \sin{d_x} \sin{d_y} - \sin{d_y} \sin{d_{z}} \cos^{2}\phi -
  \sin{d_x} \sin{d_{z}} \sin^{2}\phi\\
  & \geq & \sin{d_x} \sin{d_y} \left(1 - \cos^{2}\phi - \sin^{2}\phi
  \right)\\
  & = & 0
\end{eqnarray*}
and $1-\cos\omega\geq0$, 
\begin{eqnarray*}
  f_{1} & \leq & f_{1} \left(\theta=0,\,\phi,\,\omega \right) = f_{1}
  \left( \theta = \pi,\,\phi,\,\omega\right)\\
  & = &
  \sin{d_x} \sin{d_y} + \cos\omega \left(\sin{d_x} + \sin{d_y} \right)
  \sin{d_{z}}\\
  & \equiv & f_{1max}.
\end{eqnarray*}

\subsection{Analysis of $f_2$}

From $0\leq{d_{z}}\leq{d_y}\leq{d_x}\leq\frac{\pi}{2}$, we can get
the inequality:
\[
a \geq \left\{
  \begin{array}{c}
    b\\
    c
  \end{array} 
\right\} \geq \left\{
  \begin{array}{c}
    g\\
    p
  \end{array}
\right\} \geq q.
\]

We also have the relations 
\begin{eqnarray}
  &  & \lambda + \mu + \gamma + \eta + \nu + \xi \nonumber \\
  & = & -2 \left(\cos\alpha \cos\theta + \cos[\beta - \phi] \sin\alpha
    \sin\theta \right) \sin\omega \nonumber \\
  & = & -2 \sqrt{\cos^{2}\theta + \cos^{2}[\beta - \phi]
    \sin^{2}\theta} \sin \left(\alpha + \varphi\right) \sin\omega 
  \nonumber \\
  & \leq & 2 \left(\sqrt{1 + \left(\cos^{2}[\beta - \phi] -1 \right)
      \sin^{2}[\theta]} \right) \sin\omega \nonumber \\
  & \leq & 2\sin\omega \label{eq:full}\\
  & \leq & 2,\nonumber 
\end{eqnarray}
where
$\tan\varphi=\frac{1}{\cos\left(\beta-\phi\right)\tan\left(\theta\right)}$.
The equality holds when $\beta=\phi$, $\alpha-\theta=\pm\pi$,
$\omega=\frac{\pi}{2}$ or $\beta-\phi=\pm\pi$, $\alpha+\theta=\pi$,
$\omega=\frac{\pi}{2}$.

\begin{eqnarray}
  &  & \lambda + \mu + \gamma + \eta \nonumber \\
  & = & 2 (\cos\alpha \cos\theta + \sin\alpha \sin\beta \sin\theta
  \sin\phi) \sin\omega \nonumber \\
  & \leq & 2 \sin\omega \label{eq:nonfull}\\
  & \leq & 2,\nonumber 
\end{eqnarray}
\begin{equation}
  \lambda + \gamma = T^{z} S_{x}^{y} -T^{y} S_{x}^{z} =
  \left(\vec{S_{x}} \times \vec{T}\right)_{x} \leq 1,
\end{equation}

\begin{equation}
  \nu + \mu = T^{x} S_{y}^{z} - T^{z} S_{y}^{x} = \left(\vec{S_{y}}
    \times \vec{T}\right)_{y} \leq 1,
\end{equation}

\begin{proposition}
\label{prop:proposition} $f_2\le a+b$.
\end{proposition}

To see this, we consider
\begin{eqnarray*}
  a + b - f_2 & = & a + b- \left(\lambda a + \mu b + \gamma c + \eta g +
    \nu p + \xi q \right)
\end{eqnarray*}
1) If $-\eta{g}-\xi{q}\geq{0}$,
\begin{eqnarray*}
  &  & a + b - f_2\\
  & = & \left(1 - \lambda \right) a + \left(1 - \mu\right) b - \gamma c
  - \nu p - \eta g - \xi q\\
  & \geq & \left(1 - \lambda - \gamma \right) c + \left(1 - \mu - \nu
  \right) p - \eta g - \xi q\\
  & \geq & 0;
\end{eqnarray*}
2) if $-\eta{g}-\xi{q}\leq{0}$, 
\begin{eqnarray*}
  &  & -\eta g-\xi q \leq 0\\
  & \Leftrightarrow & \left(- \eta \cos{d_y} - \xi \cos{d_x} \right)
  \sin{d_{z}} \leq 0\\
  & \Leftrightarrow & \left(- \eta \cos{d_y} - \xi \cos{d_x} \right)
  \sin{d_y} \leq \left(- \eta \cos{d_y} - \xi \cos{d_x} \right) 
  \sin{d_{z}}\\
  & \Leftrightarrow & -\eta h-\xi p \le - \eta g - \xi q
\end{eqnarray*}
 where $h=\cos{d_y}\sin{d_y}$. 

We also have the relation:
\begin{displaymath}
  a \geq \left\{
    \begin{array}{c} 
      b\\c
    \end{array} 
  \right\} \geq h \geq p \geq 0.
\end{displaymath}

So we get
\begin{eqnarray*}
  &  & a + b - f_2\\
  & = & \left(1 - \lambda\right) a + \left(1 - \mu\right) b - \gamma c
  - \nu p - \eta g - \xi q\\
  & \geq & \left(1 - \lambda - \gamma\right) c + \left(1 - \mu\right) b
  - \nu p - \eta h - \xi p\\
  & \geq & \left(2 - \lambda - \gamma - \mu - \eta\right) h - \nu p -
  \xi p\\
  & \geq & \left(2 - \lambda - \gamma - \mu - \eta - \nu 
    - \xi \right) p\\
  & \geq & 0.
\end{eqnarray*}
Therefore we conclude from above that $a+b\geq{f_2}$, where the
equality is satisfied iff
$\theta=0,\pi$, $\alpha=\pi$, $\omega=\frac{\pi}{2}$ (that is
$T^{z}=-1$). 

Let $f'=f{}_{1max}+f_{2}\geq{f}$. Because
$f{}_{1max}(\omega_{1})<f{}_{1max}\left(\omega_{2}\right)$ when
$0\leq\omega_{2}<\omega_{1}\leq\pi,$ and $f_{2}\leq
f\left(\theta=0,\pi\,\omega=\frac{\pi}{2}\right)$, we can get the
conclusion that $f$ arrives at its maximum in the domain
$0\leq\omega\leq\pi/2$.

\begin{proposition}
$f_2\leq\left(a+b\right)\sin\omega$, when
$0\leq\omega\leq\frac{\pi}{2}$.
\end{proposition}

From Eq.(\ref{eq:full}) and Eq.(\ref{eq:nonfull}), we have the
inequality:
\[
\lambda + \mu + \gamma + \eta + \nu + \xi \leq 2 \sin\omega,\text{ if } 0
\leq\omega \leq \frac {\pi} {2}
\]
 and
\[
\lambda + \mu + \gamma + \eta \leq 2 \sin\omega, \text{ if } 0 \leq
\omega \leq \frac {\pi} {2}.
\]
Also we have the relations:
\begin{eqnarray*}
  \eta + \xi & = & T^{y} S_{z}^{x} - T^{x} S_{z}^{y}\\
  & = & 2 \sin \frac {\omega} {2} \sin\alpha \sin\theta \left[\cos
    \frac {\omega} {2} \cos\left(\phi - \beta\right) - \sin \frac
    {\omega} {2} \cos\theta \sin\left(\phi - \beta\right) \right]\\
  & \le & 2 \sin \frac {\omega} {2} \sin\theta \sqrt{\cos^{2} \frac
    {\omega} {2} + \sin^{2} \frac {\omega} {2} \cos^{2} \theta}\\
  & = & 2 \sin \frac {\omega} {2} \sqrt{- \left(\sin \frac {\omega} {2}
      \sin^{2}\theta - \frac {1} {2 \sin \frac {\omega} {2}}
    \right)^{2} + \left(\frac {1} {2 \sin \frac {\omega} {2}}\right)^{2}}
\end{eqnarray*}
Then if $\sin^{2}\theta=\frac{1}{2\sin^{2}\frac{\omega}{2}}\le1$,
then $\eta+\xi\le1$. If $\sin^{2}\frac{\omega}{2}\le\frac{1}{2}$,
i.e., $\omega\le\frac{\pi}{2}$, then $\eta+\xi\le\sin\omega$.

Similarly, when $0\leq\omega\leq\frac{\pi}{2}$, we have
\begin{eqnarray*}
  \lambda + \gamma  & \leq & \sin\omega, \\
  \nu + \mu & \leq & \sin \omega, 
\end{eqnarray*}

Therefore we can prove this proposition using the method used in the
proof of proposition (\ref{prop:proposition}) just replacing 1 and 2
with $\sin\omega$ and $2\sin\omega$ respectively.

\subsection{Conclusion}

Therefore 
\begin{eqnarray*}
  f & \leq & f' \leq f{}_{1max} + \left(a + b\right) \sin\omega\\
  & = & \sin{d_x} \sin{d_y} + \left(\sin{d_x} + \sin{d_y}\right)
  \sin\left(\omega + {d_{z}}\right)\\
  & \le & \sin{d_x} \sin{d_y} + \sin{d_x} + \sin{d_y},
\end{eqnarray*}
the equality holds when $\alpha=k\pi$, $\theta=\left(k+1\right)\pi$,
and $\omega=\frac{\pi}{2}-{d_{z}}$.

\end{appendix}

\bibliography{/Users/zhou/Documents/Work/QST/ref/QSTP}

\end{document}